\documentclass[pra,twocolumn,citeautoscript, showpacs]{revtex4}

\usepackage{graphicx}
\usepackage{dcolumn,comment}

\begin{document}

\title{Controlling Metamaterial Resonances with Light}
\author{Sangeeta Chakrabarti, S. Anantha Ramakrishna and Harshawardhan Wanare}
\affiliation{Department of Physics, Indian Institute of Technology Kanpur 
208016, India}

\begin{abstract}

We investigate the use of coherent optical fields as a means of 
dynamically controlling the resonant behaviour of a variety of composite 
metamaterials, wherein the metamaterial structures are embedded in a 
dispersive dielectric medium. Control and switching is implemented by 
coherently driving the resonant permittivity of the embedding medium by 
applied optical radiation. The effect of embedding Split ring resonators 
(SRR)  in a frequency- dispersive medium with Lorentz-like dispersion or 
with dispersion engineered by electromagnetic induced transparency (EIT), 
is manifested in the splitting of the negative permeability band, the 
modified (frequency-dependent) filling fractions and dissipation factors. 
The modified material parameters are strongly linked to the resonant 
frequencies of the medium, while for an embedding medium exhibiting EIT, 
also to the strength and detuning of the control field. The robustness of 
control against the deleterious influence of dissipation associated with 
the metallic structures as well as the inhomogeneous broadening due to 
structural imperfections is demonstrated. Studies on plasmonic 
metamaterials that consist of metallic nanorods arranged in loops and 
exhibit a collective magnetic response at optical frequencies are 
presented. Control and switching in this class of plasmonic nanorod 
metamaterials is shown to be possible, for example, by embedding these 
arrays in a Raman active liquid like CS$_2$ and utilizing the Inverse 
Raman Effect.

\end{abstract}

\maketitle

In recent times, several new developments have considerably altered our 
perception of the interaction of electromagnetic waves with material 
media in general and light-matter interaction in particular. 
On one hand, the coherent control of quantum systems utilizing the quantum 
interference route~\cite{harris_Phystoday} has enabled researchers to demonstrate 
superluminal~\cite{wang_Nature2000} and subluminal~\cite{hau_Nature1999} 
propagation of light, stopped light~\cite{lukin, stopped_light}, render an 
opaque medium transparent at a designated frequency (Electromagnetically 
Induced Transparency or EIT)~\cite{harris_PRL1990,harris_PRL1991} and 
generate an ultra-high index of refraction~\cite{scully_PRL1991},  
while on the other, the development of structured composite media 
known as metamaterials~\cite{sar_rop} has opened up new possibilities like 
negative refraction~\cite{smith_PRL2000,shelby_Science2001}, perfect lenses
~\cite{pendry_PRL2000}, "invisibility cloaks"~\cite{pendry_Science2006}, etc.
These effects have been realised using on two different approaches to the study of 
light-matter interaction. While coherent control is a purely quantum mechanical phenomenon,
 metamaterials can be analysed within the framework of classical electromagnetic theory.

One of the most attractive features of metamaterials, (whose resonant response is governed 
by their underlying geometric structure) is the fact that they can be designed to operate at 
any predetermined frequency. In addition, it is also possible to modify and control their 
response. However, the very resonant nature of the response of metamaterials 
that gives rise to their novel properties also hinders their applicability 
owing to the high levels of dissipation associated with their response. 
Although designing metamaterials that function at optical frequencies is 
difficult, it has been demonstrated by several groups~\cite{shalaev_NaturePhot2007}. 
Another drawback of metamaterials arises from the fact that since their response depends 
on their geometrical structure, their properties are fixed once they have 
been fabricated. However, to incorporate greater flexibility to their structure
and performance, metamaterials need to be both reconfigurable as well as controllable.
Recently, control of metamaterial response has been demonstrated, both by passive as well as 
dynamical methods~\cite{padilla,lakhtakia,liqcrys,sar_prb}. 

In this paper, we demonstrate the possibility of controlling the magnetic 
response of metamaterials consisting of Split Ring Resonators (SRR)~\cite
{pendry_MTT1999} by actively tuning the capacitance of these structures at 
near-infra-red and optical frequencies. We have used the term metamaterial to signify 
the bare metamaterial while the term `composite metamaterials' indicates metamaterials 
embedded in a dispersive background. Our approach to this scheme is both simple and 
straightforward. It makes use of quantum mechanical phenomena 
to parametrically control a purely classical effect. This is achieved by embedding the 
metamaterial within a coherent atomic/molecular medium. An appropriate choice of the 
resonance frequencies of the metamaterial and the embedding medium, together 
with the control field and its detuning (in the case of EIT) results in 
complete control of the effective magnetic response of the metamaterial.
This occurs through the interaction of the narrow atomic resonance of the background 
dielectric with the broader magnetic resonance of the metamaterial. Along with the control 
over the magnetic response, it is seen that the dissipation inherent in the metamaterial 
can be reduced significantly. This scheme may be used to produce metamaterials which are 
reconfigurable as well as controllable by choosing a slightly different resonance frequency 
for the embedding medium. 
In addition, we also show how an array of nanorods that exhibits a magnetic response 
which is plasmonic in origin~\cite{engheta_oe} can also be controlled via the 
self-capacitance of the array. Further, the same array of nanorods 
(which is known to behave like an effective plasma when radiation whose electric field is axially oriented
is incident upon it), can be made transmittive 
at frequencies well below its effective plasma frequency via the resonantly 
enhanced permittivity of its background medium.

\begin{figure}[tbp]
\includegraphics[ angle=-0,width=1.0\columnwidth] {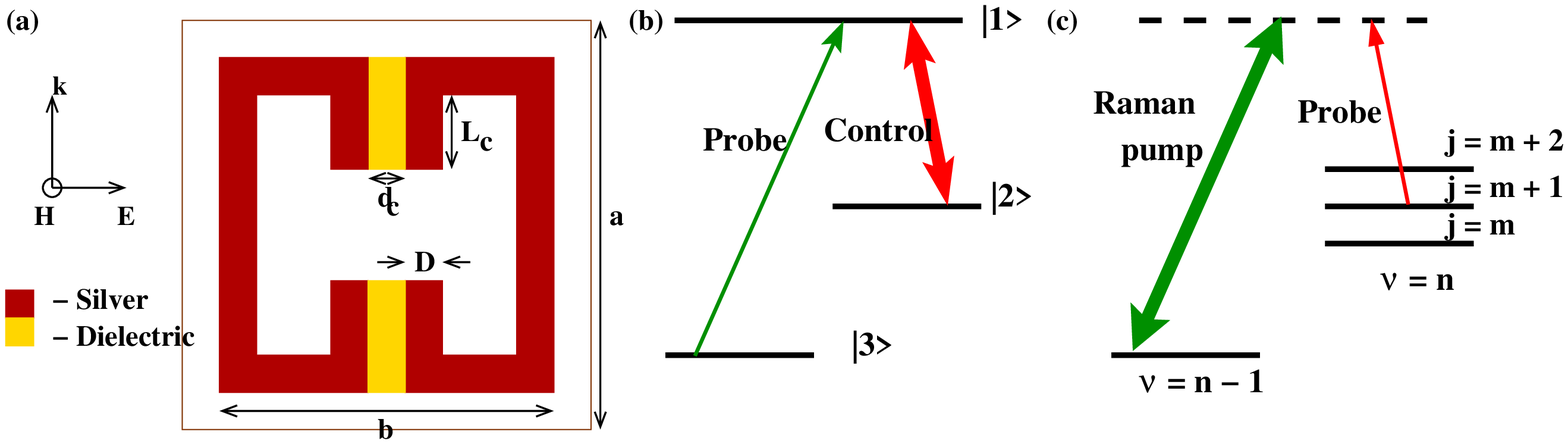}
\caption{ Atomic level schemes for realizing (a) Lorentz permittivity 
dispersion(by means of a Raman transition) and (b) EIT-like permittivity dispersion.\\
(c) The homogenizable SRR~\cite{obrien_jpcm2002} used in our calculations. 
A slab consisting of these SRRs exhibits negative magnetic permeability
in a frequency range above the magnetic resonance frequency
$\omega_m = 1/\sqrt{LC}$. The following structural parameters
are used in our calculations: a = 600nm, b = 312nm,
L = 144nm, d = 24nm, D = 24nm. } \label{level_schemes}
\end {figure}

\section{Analytical description}

Let us consider a metamaterial consisting of an array of symmetric split 
ring resonators (SRR), as shown in Fig.~\ref{level_schemes}(a).These arrays lie in 
the horizontal plane and are stacked up vertically. The size of the individual 
inclusions ($\sim$ $\lambda$/10) makes it possible to describe the array as an 
effective medium at the frequencies of interest. This medium has been shown to possess 
a resonant magnetic response and its effective permeability is $\mu_\mathrm{bare}$ 
given by~\cite{obrien_jpcm2002}:
\begin{equation}
\mu_\mathrm{bare}(\omega) = 1 + 
\frac{f_m \omega^2}{\omega_m^2 - \omega^2 -i\Gamma_m\omega}.
\end{equation}
Here, $\omega_m = \frac{1}{\sqrt {LC}}$ where $L$ and $C$ are the total 
inductance and the total capacitance of the structure~\cite{obrien_jpcm2002}, 
$f_m$ is the filling fraction while the factor $\Gamma_m$, which governs 
the effective dissipation present in the SRR medium, is related to the intrinsic 
dissipation rate of the metal defined according to the Drude model for the dielectric 
permittivity. This medium exhibits a negative permeability leading to a band gap 
in a frequency range above $\omega_m$ if $\Gamma_m$ is small enough.
We study the effect of embedding an array of such SRRs in a  medium having a 
frequency-dispersive dielectric permittivity. This may either be a resonant
 Lorentz-type dispersion, or under the application of a pair of coherent 
control and probe fields, EIT-like. In the first case, the relative 
permittivity of the medium is given by the expression

\begin{equation}
\epsilon(\omega) =  1 + \frac{f_e^2}
{\omega_e^2 - \omega^2 - i\gamma_e \omega}, 
\end{equation}
where $f_e^2 = Ne^2/m \epsilon_0$ (N is the density of (atomic) oscillators, 
$\epsilon_0$ is the permittivity of free space while $e$ and $m$ are the
electronic charge and  mass, respectively), $\omega_e$ is the dielectric 
resonance frequency and $\gamma_e$ is the dissipation factor. Such a 
dispersion can be realized by using for example, resonant quantum dots or choosing a 
Raman transition~\cite{kalugin_OL2006} with a strong pump field ensuring that 
the resonant Raman probe frequency is close to the magnetic resonance 
frequency of the metamaterial, $\omega_m$.

In the second case, when the medium exhibits EIT, the permittivity for the 
probe field is given by
~\cite{scully_book}

\begin{equation}
\epsilon_{_\mathrm{EIT}}(\omega) =  1 + 
\frac{\kappa (\omega_\mathrm{1} - \omega)}{(\omega_\mathrm{1} - \omega)^2 - 
\frac{\Omega_c^2}{4} - i \gamma_\mathrm{13} (\omega_\mathrm{1} - \omega) }.
\end{equation}
Here $\omega_\mathrm{1} = \frac{E_\mathrm{1} - E_\mathrm{3}}{\hbar} $, 
$\Omega_c$ is the Rabi frequency of the control field defined as 
$\Omega_c = \frac{\vec d_{12} \cdot \vec{E}}{\hbar}$, 
$\kappa = (N_a \vert \langle d_{13} \rangle \vert^2)/(\epsilon_0 \hbar)$ is the 
strength of the transition, $N_a$ is the atomic density, $\gamma_\mathrm{13}$  and $d_{13}$ 
represent the decay rate and the dipole moment,  respectively, between the atomic 
levels $\left| 1 \right \rangle$ and $\left| 3 \right \rangle$ while $d_{12}$ represents
the dipole moment between levels $\left| 1 \right \rangle$ and 
$\left| 2 \right \rangle$(see Fig.~\ref{level_schemes}(b)).
It should be noted that in the absence of the control field, the dispersion 
obtained is the same as in the case of the Lorentz permittivity.

In the presence of the embedding medium, the capacitance of individual 
SRR units becomes strongly frequency-dependent. This makes the resonance
frequency of the composite medium ($\omega_\mathrm{eff}$) appear frequency
dispersive as:
 $\omega_\mathrm{eff} = \frac{\omega_m}{\sqrt{\epsilon(\omega)}}$.

The sharply resonant nature of $\epsilon (\omega)$ results in the SRR medium showing 
a resonant magnetic response at multiple frequencies. These predominantly occur above 
and below $\omega_m$, depending on whether the frequency of the incident wave is 
greater than or lesser than $\omega_e$.  The new resonance frequencies are the 
solutions of the transcendental equation

\begin{equation}
\mathrm{Re}\left[\omega_m^2/\epsilon(\omega)\right]- \omega^2 =  0, 
\label{res_condn}
\end{equation}
while Im[$\omega_m^2/\epsilon(\omega)$] quantifies the dissipation in the 
effective medium. Essentially, the presence of the dispersive dielectric 
permittivity of the embedding medium turns the SRR into a resonantly driven, 
actively tuned LC-circuit. The capacitance of the of the LC-circuit becomes
frequency dependent, allowing the satisfaction of the resonance conditions 
Eq. (\ref{res_condn}) at multiple frequencies with the corresponding 
negative-$\mu$ band gap of the `bare' SRR splitting into two or more band gaps.

The filling fraction $f_\mathrm{eff}$ and the dissipation parameter 
of the composite metamaterial can be viewed as effective frequency dependent quantities 
whose behaviours are greatly influenced by the resonant background. The 
effective filling fraction $f_\mathrm{eff}$ and the effective dissipation 
parameter $\Gamma_\mathrm{eff}$ have the following generic forms in the 
presence of a frequency-dispersive background:

\begin{eqnarray}
 f_\mathrm{eff}(\omega) = f_m \left[\{ \mathrm{Real}
\left[\epsilon (\omega)\right] \}^2 
+ \{ \mathrm{Im}\left[\epsilon (\omega) \right] \}^2 \right],  \\
\Gamma_\mathrm{eff} (\omega) = \omega_m^2 
\frac{\mathrm{Im}\left[\epsilon (\omega) \right]}{\omega} 
+ \Gamma_m \{ \mathrm{Real}\left[\epsilon (\omega) \right] \}^2  \nonumber \\ 
+ \Gamma_m\{ \mathrm{Im}\left[\epsilon (\omega) \right] \}^2 . 
\end{eqnarray}

\begin{figure}[t]
\includegraphics[ angle=-0,width= 1.0\columnwidth] {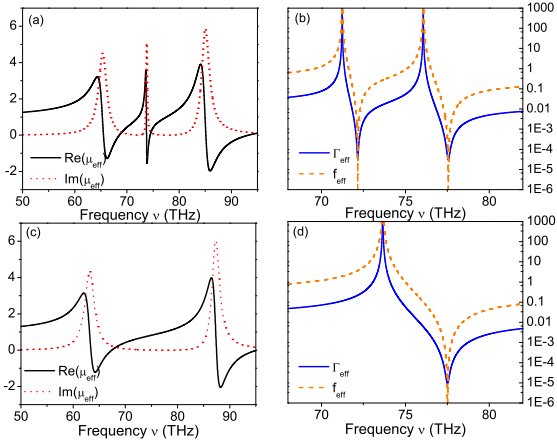}
\caption{(a)The frequency dependent $\mu_\mathrm{eff}$ and (b)
$\Gamma_\mathrm{eff}$ and filling fraction $f_\mathrm{eff}$ 
(obtained analytically) for the SRR metamaterial in the presence 
of a embedding dielectric medium displaying EIT.
(c) and (d): The same quantities for the SRR metamaterial embedded 
in a medium which has a resonant Lorentz permittivity.} \label{parameter_plots}
\end {figure}

Let us now study the effect of the Lorentz and the EIT dispersion 
on the properties of the metamaterial in some detail. If the background dielectric medium
has a Lorentz-type dispersion, the following expressions for $f_\mathrm{eff}$ and 
$\Gamma_\mathrm{eff}$ are obtained:

\begin{eqnarray}
f_\mathrm{eff} = f_m \left[1 + \frac{f_e^2(2(\omega_e^2 - \omega^2) + f_e^2)}
{(\omega_e^2 - \omega^2)^2 + (\gamma_e \omega)^2}\right],\\
\Gamma_\mathrm{eff} = \Gamma_m + 
\frac {\Gamma_m f_e^2 (2(\omega_e^2 - \omega^2) + f_e^2) +  
\omega_m^2 f_e^2 \gamma_e}{(\omega_e^2 - \omega^2)^2 + (\gamma_e \omega)^2}. 
\end{eqnarray}
Along with the two new magnetic resonance frequencies and a split in 
the negative $\mu$-band, a region of significantly reduced absorption is 
observed between the two resonances. In fact, the Im[$\mu_\mathrm{eff}$] is 
observed to pass through a minimum at the resonant frequency of the bare 
SRR if $\omega_m = \omega_e$. There is also a significant change in the 
nature of $f_\mathrm{eff}$ and $\Gamma_\mathrm{eff}$ around the frequency 
$\omega_e$. The large values of $f_\mathrm{eff}$ at $\omega_e$ is due to the resonant
enhancement of the cross-section of the SRR structural unit: the SRR appear much larger
to radiation at resonance than they actually are. The reduction in absorption at $\omega_e$ 
is due to the vanishing of the currents in the SRR. This results from an interplay of two 
currents giving rise to different charge densities. The first of these is due to the 
current induced in the SRR unit by the magnetic field of the incident radiation. 
The second arises due to electric polarization of the embedding dielectric background 
by the electric field of the incident light. At resonance, the two are out of phase 
with each other leading to the freezing of currents in the SRR. The proximity of 
$\omega_e$ to $\omega_m$ determines the extent of modulation of the $\mu_\mathrm{eff}$ 
response. In other words, the interaction between the electric and magnetic 
resonances is the strongest when $\omega_e$ lies in the vicinity of $\omega_m$, 
($\omega_m \pm \Gamma_m$). Otherwise, the two resonances are virtually uncoupled.  
 
The magnetic response of the composite metamaterial embedded in an EIT medium
exhibits multiple magnetic resonances (Fig. \ref{parameter_plots}(b)). This is again
accompanied by drastically reduced absorption within two frequency bands. The strength and 
the detuning of the control field can both be used to shift the positions, widths, slopes 
and dissipation associated with the various resonances in $\mu_\mathrm{eff}$. 
In fact, this scheme is more amenable to control as compared to the one described earlier 
because  the detunings of the probe and the control beams (both positive and 
negative) can be used to shift the EIT line centre and hence, the frequencies 
at which the absorption in the metamaterial is reduced. Again, the effective 
filling fraction and the modified dissipation  parameter are given by 

\begin{eqnarray}
f_\mathrm{eff} = f_m \left[1 + \frac{\kappa \Delta(2(\Delta^2 - 
\Omega_\mathrm{c}^2) + \kappa \Delta)}
{(\Delta^2 - \Omega_\mathrm{c}^2/4)^2 + (\gamma_\mathrm{12} \Delta)^2}\right],\\
\Gamma_\mathrm{eff} = \Gamma_m + 
\frac {\Gamma_m \kappa \Delta (2(\Delta^2 - \Omega_\mathrm{c}^2) + \kappa 
\Delta) +  \omega_m^2 \kappa \Delta \gamma_\mathrm{12}}{(\Delta^2 - 
\Omega_\mathrm{c}^2)^2 + (\gamma_\mathrm{12} \Delta)^2}, 
\end{eqnarray}
where $\Delta = \omega_1 - \omega$.

From the preceding expressions, we find that the new filling fraction 
$f_\mathrm{eff}$ increases close to the resonant frequency while 
$\Gamma_\mathrm{eff}$ exhibits sub-natural $(\Gamma_\mathrm{eff} \ll \Gamma_m)$ 
values around the atomic resonances (Fig.~\ref{parameter_plots}(c)). At the EIT 
line centre frequency ($\omega_1$), $f_\mathrm{eff}$ and $\Gamma_\mathrm{eff}$ are
 found to have the same values as the bare SRR medium. As before, the 
reduction in absorption is a manifestation of the vanishing currents in the 
SRR loops. It is interesting to note that if losses are reduced significantly,
the regions where $\mathrm{Re}\left[ \mu_\mathrm{eff}\right] < 0$ gradually disappear,
(see Fig. ~\ref{parameter_plots}), in agreement with the findings of 
Stockman~\cite{stockman_prl}.

If the metamaterial has been designed to exhibit a resonant response at the 
frequency of the EIT line centre in the absence of medium, the resonant 
magnetic response of the composite metamaterial is found to be such that $\mathrm{Re}
\left[ \mu_\mathrm{eff}\right] \rightarrow 1$ (with an associated imaginary 
part) exactly at the EIT line centre. However, if $\omega_m \neq \omega_1$, the response 
of the composite at the EIT line centre is simply the value of $\mu_\mathrm{bare}$ 
at the frequency $\omega_\mathrm{1}$. It should be emphasized that since we are dealing 
with a resonant system with a highly dispersive lineshape, changes of the control parameter
($\epsilon_L$ or $\epsilon_{EIT}$) can lead to large-scale changes in response 
($\mu_\mathrm{eff}$) at a given frequency, due to the shift in the resonance frequency.

\section{Numerical Simulation of the SRR metamaterial at mid-IR frequencies}

Numerical simulations using the PHOTON codes based on the Transfer Matrix 
Method~\cite{mackinnon_prl92,pendry_jmo94} were carried out to verify our 
theoretical predictions. We present the canonical example of subwavelength-sized SRR 
made of silver at mid-infrared (mid-IR) frequencies. The metamaterial shows magnetic 
activity for applied radiation polarized with the magnetic field along the SRR axis. 
The structures were modelled as being embedded in a resonant background. We have used 
experimentally obtained parameters for silver~\cite{johnson_christy} in our calculations. 
(We use silver for all metallic inclusions in this paper.)

\begin{figure}[tbp]
\includegraphics[angle=-0,width=1.0\columnwidth] {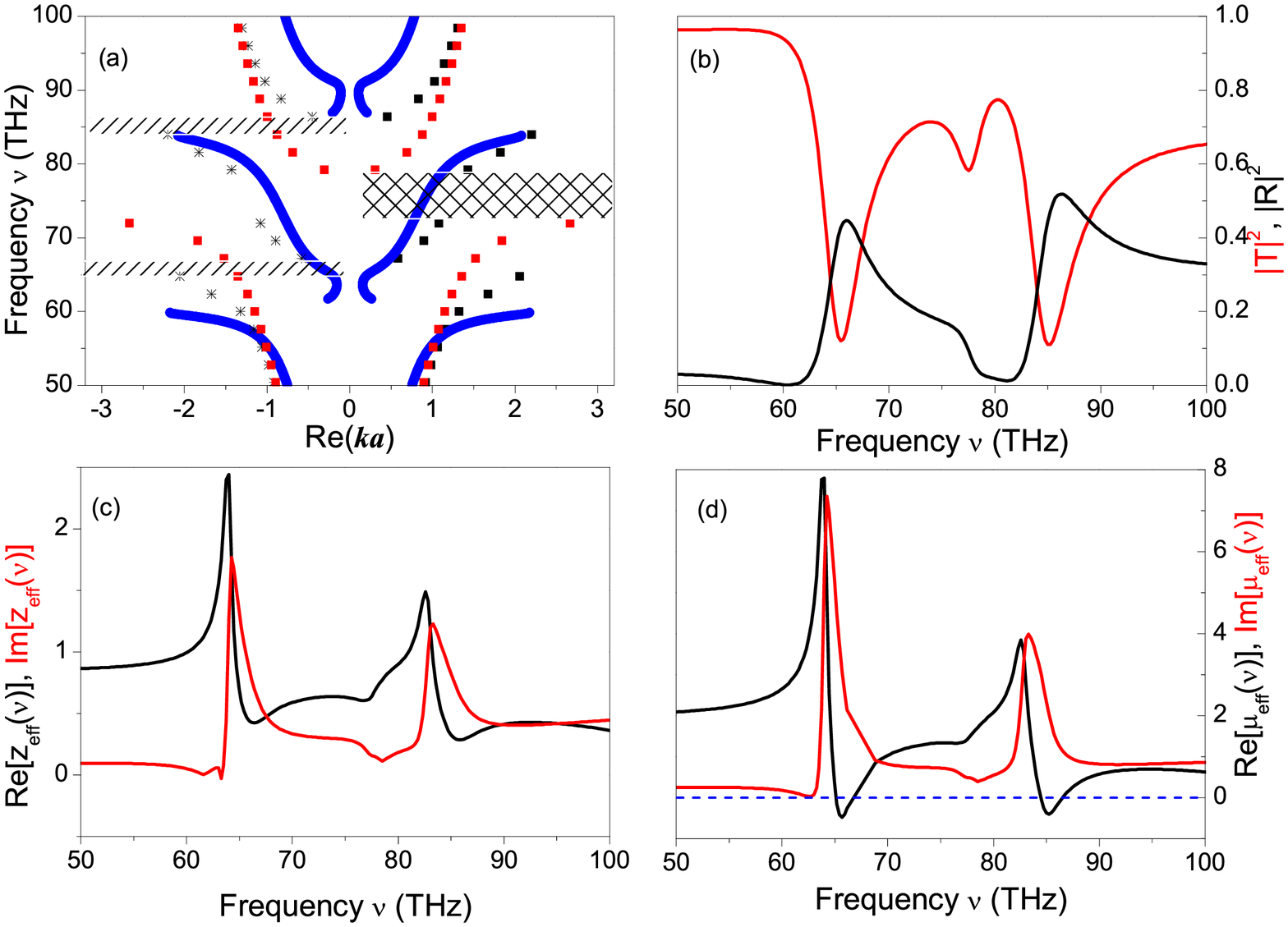}
\caption{(a): The computed band structures for the bare SRR (red circles) and 
the composite metamaterial embedded in a Raman medium (black squares) under 
the condition $\omega_e = \omega_m$. The blue lines represent the dispersion 
predicted by the analytic formula. The band gaps due to $\mu_\mathrm{eff} < 0$ 
are indicated by the cross-hatched region on the right 
(for the bare SRR whose $\omega_m$ = 74.9THz) and the the two hatched regions 
on the the left (for the band gaps resulting from the inclusion of the 
resonant Raman medium). \\
(b) The reflectivity and the transmittivity for a calculated for a slab 
consisting of a single layer of SRR. Note the presence of two stop bands 
(low $|T|^2$) corresponding to the band gap frequencies in (a)
and the new transmittive band that develops within the original band gap.\\
(c)The effective impedance Z($\nu$)of the composite metamaterial showing the 
presence of two resonances.\\
(d): The $\mu_\mathrm{eff}$, for a slab of the SRR metamaterial with an 
embedded resonant medium, retrieved using the computed reflection and the 
transmission coefficients. Note the existence of two frequency ranges where 
$\mu_\mathrm{eff} < 0$. The grey dotted line represents the zero level.} 
\label{retrieval_plots}
\end {figure}

The cross-section of the (cylindrical) SRR that we considered in this case is 
shown in Fig.~\ref{level_schemes}. Seventy-five grid points along each 
direction were used for computing the band structure as well as the 
reflection and the transmission coefficients. Invariance along the cylindrical axis 
is assumed. The bare SRR in this case has a magnetic resonance frequency of 74.9 THz. 
This metamaterial containing a Lorentz-type dielectric within its 
capacitive gaps is modelled using the following parameters: 
$f_e = 24.1$ THz, $\omega_e = 74.9$ THz and $\gamma_e = 2.4 $THz. 

As predicted in the preceding section, the negative $\mu$ band splits in the 
presence of the embedding dielectric medium. The frequencies at which the 
new band gaps are formed are consistent with the predicted values. The mismatch 
between the results of the TMM calculation and the analytic formula can be 
attributed to the presence of parasitic capacitances in the SRR. These 
capacitances, which were neglected in the theoretical model used to describe 
the functioning of the SRR~\cite{obrien_jpcm2002}, are automatically taken 
into account in the TMM simulation. 

The results of the band structure calculation were compared with 
the transmittivity and the reflectivity calculated for a slab consisting of
 a single layer of the SRR. A new propagating band (indicated by 
a peak in the transmittivity) is found to occur within the stop band region of 
the bare SRR. The frequency intervals corresponding to the new band gaps 
(regions of high reflectivity and low transmittivity) were found 
to be consistent with the band structure calculations. To illustrate the splitting 
of the negative permeability band, the effective material parameters for a 
slab consisting of four layers of the SRR-dielectric composite were determined using 
a retrieval procedure that utilizes the complex reflection and transmission coefficients
~\cite{smith_prb}. The splitting of the negative $\mu$ band is found to be 
consistent with the predicted values (Fig. \ref{retrieval_plots} (d)) and also 
with the simulated band structure and the reflectivity and the transmittivity. 
In addition, Fig. \ref{retrieval_plots}(c) shows the effective impedance 
Z($\nu$) of the metamaterial, with two resonances at the same frequencies 
as the negative permeability bands, in contrast with the single peak 
in Z($\nu$) observed for the bare SRR\cite{obrien_jpcm2002} .

\begin{figure}[h]
\includegraphics[angle=-0,width=1.0\columnwidth] {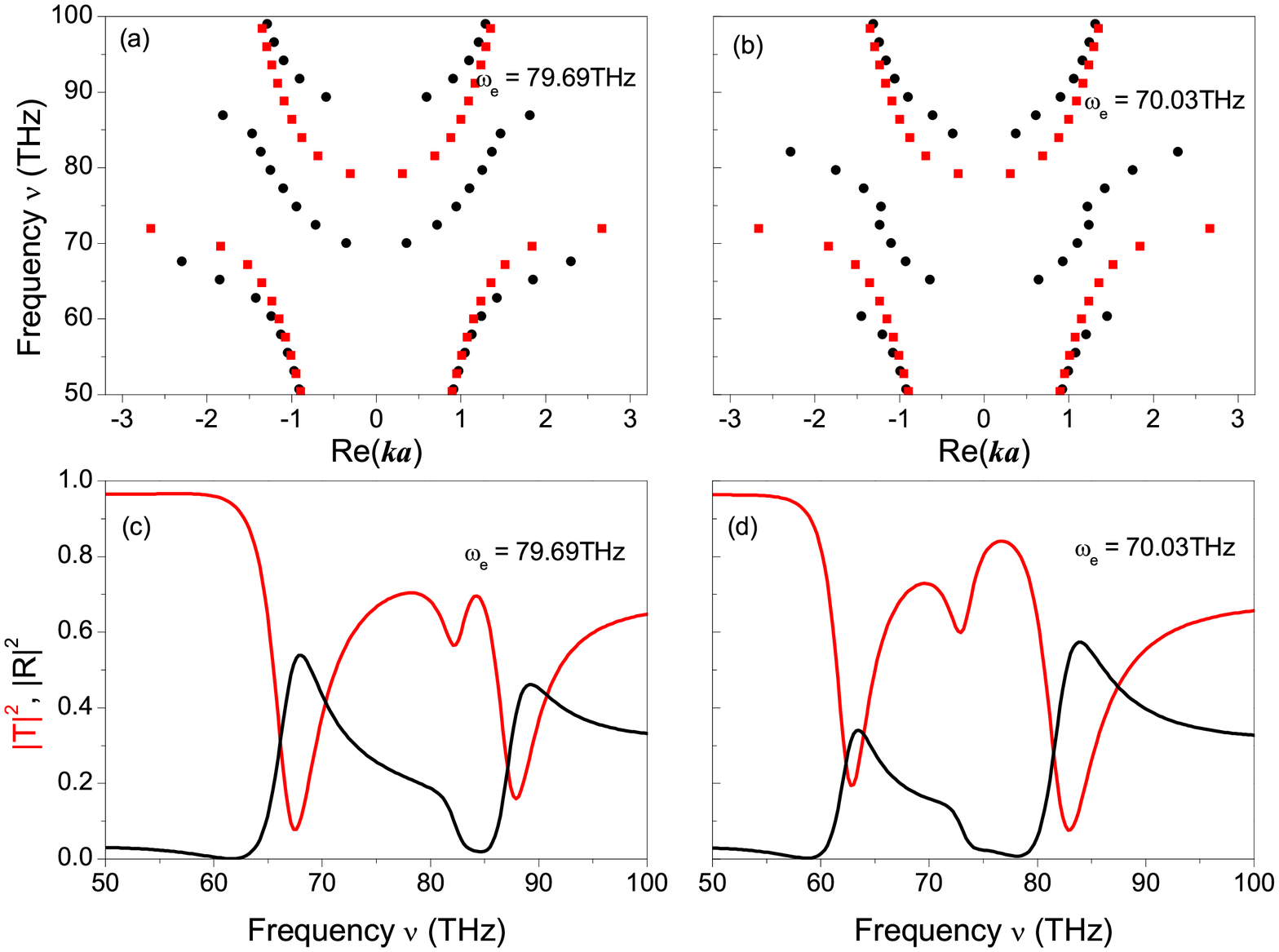}
\caption{The band structures and ($|T|^2$, $|R|^2$) for the 
composite metamaterial when $\omega_m\neq \omega_e$.\\ (a): Band structure of 
the composite metamaterial when $\omega_e$ = 79.69THz. The resulting gaps are 
blue-shifted with respect to those obtained when $\omega_e$ = $\omega_m$.\\
(b): Band structure of the composite metamaterial when $\omega_e$ = 70.03THz. 
In this case, the resulting gaps are red-shifted with respect to 
those obtained when $\omega_e$ = $\omega_m$.\\
 (c) \& (d): Reflectivity and transmittivity calculated for a slab consisting 
minima in $|T|^2$ correspond to the new band gaps.
\label{offres}} 
\end {figure}

Fig. \ref{offres} illustrates the response of the metamaterial when 
the dielectric and the magnetic resonances do not coincide, i.e., $\omega_e 
\neq \omega_m$. As before, the negative permeability band splits, but the 
new resonances are either red-shifted or blue-shifted (with respect to those 
obtained when $\omega_e = \omega_m$), depending on whether $\omega_e < 
\omega_m$ or $\omega_e > \omega_m$. A study of the reflection and 
transmission coefficients for a slab consisting of four layers of the SRR 
further confirms this behaviour. As seen in the case when $\omega_m = 
\omega_e$, two regions of enhanced reflectivity (corresponding to the band 
gaps) separated by a transmittive region (corresponding to the additional 
propagating band that develops) are observed. For multiple layers, additional
 peaks due to Fabry-P\'{e}rot resonances (arising as a result of multiple 
scattering) are observed. In each case, the calculated effective impedance 
(Z($\nu$)) of the structure (not shown here) reveals the presence of two peaks, 
signifying the two magnetic resonances. However, the effect of the resonant 
Lorentz medium weakens considerably when the two resonances are far apart.

In a similar fashion, appropriately designed SRR structures, 
embedded in resonant atomic media, can be shown to behave in the 
same way at NIR/optical frequencies~\cite{oe_iitk}. Multiple layers of
such SRR can switch sharply between highly reflecting to transmitting 
states within a narrow bandwidth. These fine features in the 
$|R|^2-|T|^2$ profile arise due to the Fabry-P\'{e}rot resonances of the 
slab. Coupled with the control possible via EIT, these materials are 
potential candidates for narrow-band switching applications.

\section{Robustness of the control}

Although it is often seen that light interacting with mesoscopic metallic 
structures experience much higher dissipation than the bulk metal itself, we 
have found that our results remain unaffected even if the intrinsic dissipation 
of the metal (silver, in this case) is increased three times from its bulk value. 

There is another aspect of these structures that needs to be taken into account. 
The resonance frequency of these metamaterials depends crucially on the accuracy with which 
they can be fabricated. Thus, in practice, each metamaterial unit 
will be different from the rest, leading to a distribution of resonance 
frequencies rather than a single resonance frequency. The consequent 
broadening of the magnetic resonance could wipe out the narrow 
band effects predicted. Fortunately, we find that the effects are rather
robust against the such inhomogeneous broadening.

\begin{figure}[tbp]
\includegraphics[angle= -0, width=1.0\columnwidth]{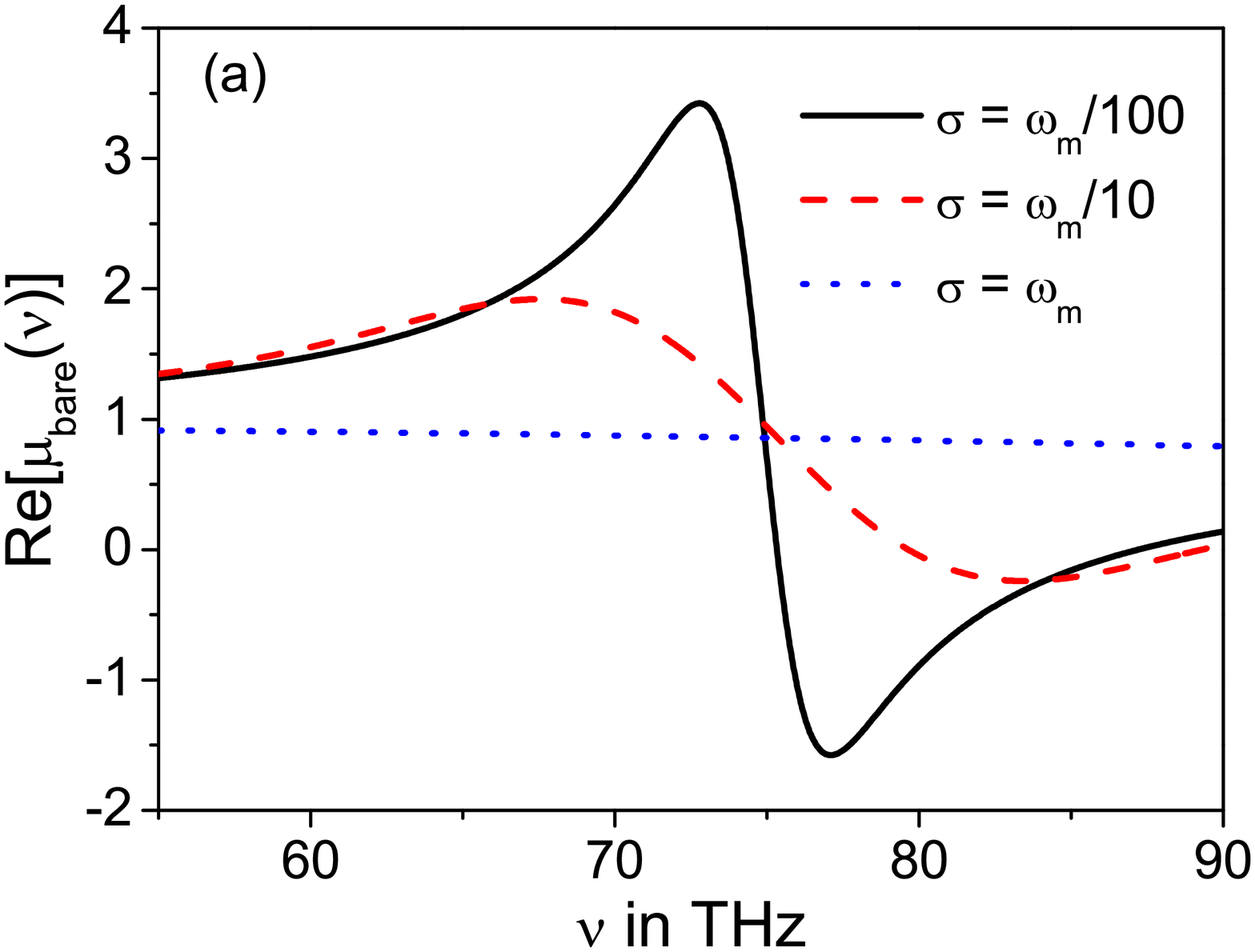}
\includegraphics[angle= -0, width=1.0\columnwidth]{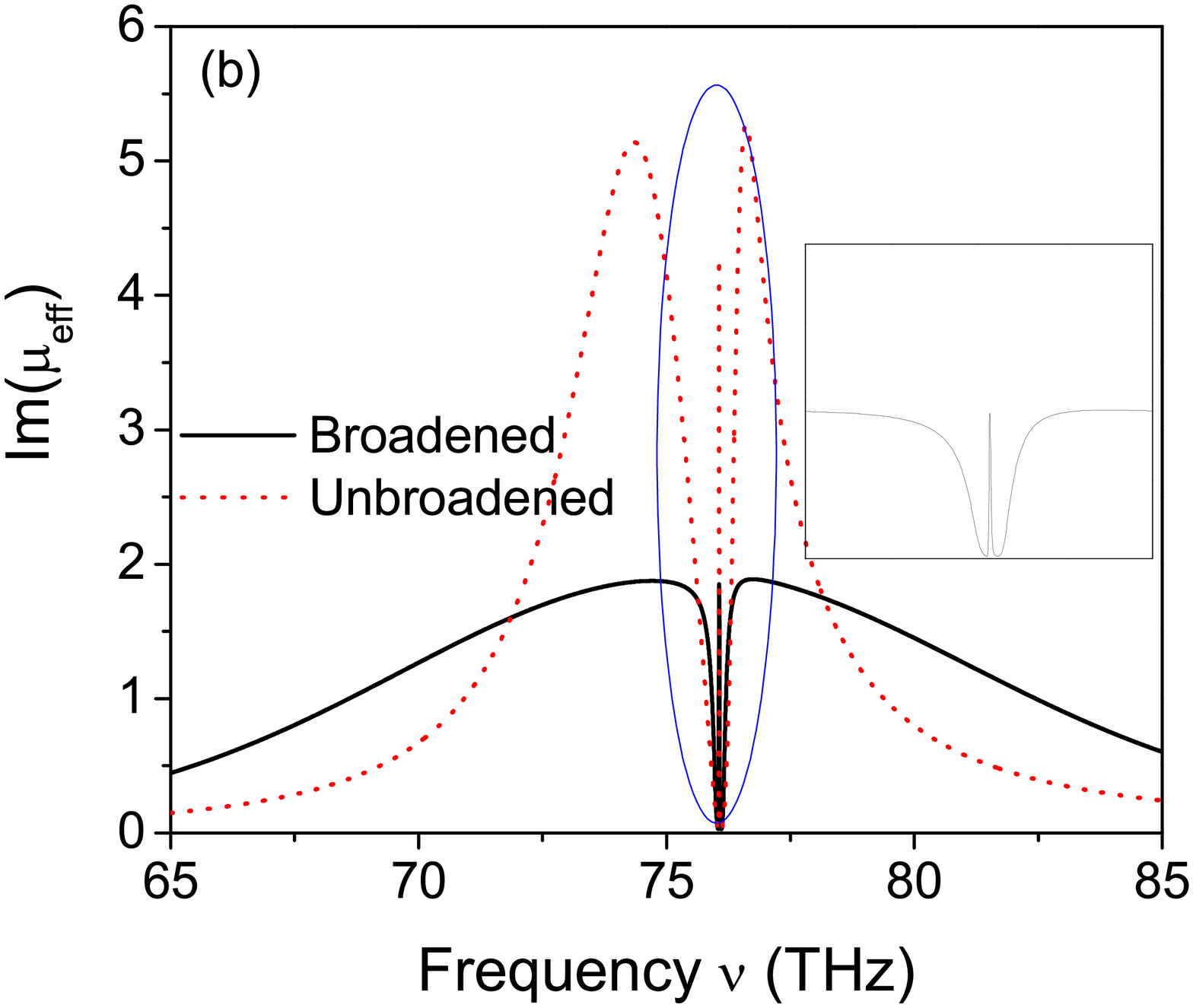}
\caption{(a)The effective permeability of the metamaterial for different
widths of the distribution of the magnetic resonance frequency. The resonant 
$\mu_{\mathrm{eff}}$ is washed out when the imperfections lead to $\sigma \sim \omega_m$. \\
(b):The broadened $\mathrm{Im}(\mu_{\mathrm{eff)}}$ of the metamaterial 
in the presence of a 
background medium which exhibits EIT, for $\sigma = \omega_m/100$. 
The inset shows the details of the 
central part of the figure where the narrow features are seen to survive 
inspite of the broadening of the overall response.\label{broadened_mu}}
\end{figure}

We consider a distribution of magnetic resonance frequencies arising due to structural 
imperfections to be normally distributed (Gaussian) about the mean frequency $\omega_m$. 
The width of this distribution ($\sigma$) is indicative of the extent of the imperfect 
structuring. Averaging over this distribution, we obtain:

\begin{eqnarray}
\langle \omega_\mathrm{eff}^2  \rangle &=& \omega_\mathrm{eff}^2 + \frac{\epsilon_R
\sigma^2}{\epsilon_R^2 + \epsilon_I^2} \\
\langle \Gamma_\mathrm{eff} \rangle &= & \Gamma_\mathrm{eff} + \frac{\epsilon_I 
\sigma^2}{\epsilon_R^2 + \epsilon_I^2}\\
\langle f_\mathrm{eff} \rangle &= & f_\mathrm{eff}
\end{eqnarray} 
Note that the effective filling fraction ($f_\mathrm{eff}$) remains unaltered.
Thus, the effective permeability of the medium is found to broaden, as shown in
Fig. \ref{broadened_mu}(a).

For values of $\sigma \sim \Gamma_m$, we find that the narrow band effects arising 
due to the presence of $\epsilon (\omega)$ survive (See Fig. \ref{broadened_mu}(b)). While 
the strength of the resonance decreases on the whole due to the larger $\Gamma_\mathrm{eff}$, 
the regions of reduced absorption between the resonances remain nearly unaltered. The 
`increased' filling fraction that remains unaffected by the inhomogeneous broadening 
compensates for the effect of the imperfections. In other words, the few resonant SRR units 
exhibit a large enough absorption cross-section such that the reduced number of 
participating SRRs due to the inhomogeneous broadening does not adversely affect the 
response on an average. Although accuracy in fabrication still remains an important issue, we 
note that the controllable effects discussed here are robust and generic to many 
metamaterials that depend upon resonances for their electromagnetic properties. 

\section{Plasmonic metamaterial at visible frequencies}

\begin{figure}[tbp]
\includegraphics[angle= -0, width=0.45\columnwidth]{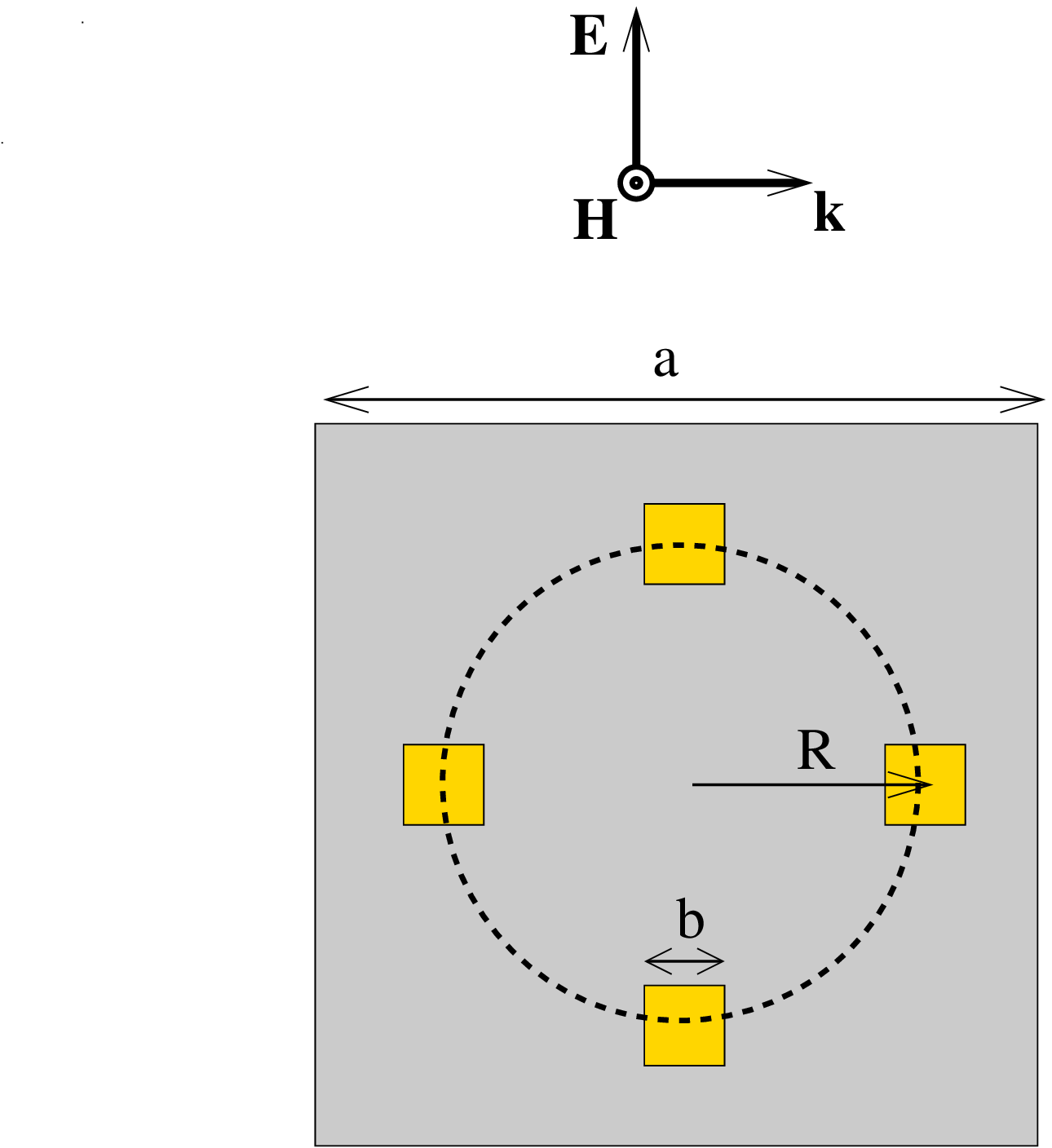}
\hspace{0.5cm}
\includegraphics[angle= -0, width=0.45\columnwidth]{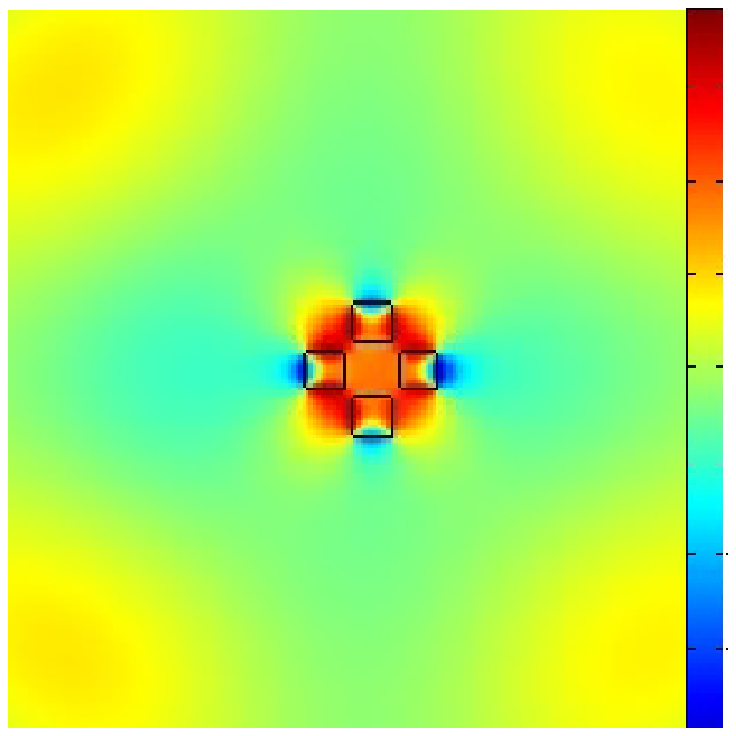}
\caption{(a): A cross-sectional view of a typical unit cell of
the plasmonic loop metamaterial. The picture shows the nanorods placed
on the circumference of a circle. The grey area represents the embedding dielectric medium.\\
(b): The magnetic response of an isolated loop inclusion
illustrating the concentration of the magnetic field inside
the loop at resonance, when excited by means of a line source.\label{unitcellp}}
\end{figure}

In this section, we present the response and control of a metamaterial consisting of 
plasmonic nanorod loop inclusions. Arrays of small resonant plasmonic spheres arranged on 
subwavelength-sized loops have been shown to exhibit a magnetic response in 
Ref.~\cite{engheta_oe}. We consider here plasmonic inclusions of metallic (silver) 
nanorods arranged on the circumference of a circle (Fig. \ref{unitcellp}(a)). The magnetic field of the incident 
radiation is assumed to be oriented along the axis of the rods. On assuming invariance along the rod axis, 
the problem becomes 
essentially two-dimensional. The loop exhibits a magnetic resonance that stems from 
an interaction driven by the individual plasmonic resonance of each nanorod. 
Each resonating nanoparticle induces a circulating {\it displacement} current around 
this loop. The electric dipole moments (perpendicular to the axis) tend to 
align along the circumference of the ring when driven at the plasmon
resonance frequency. This is confirmed by our COMSOL finite elements numerical simulations
 (Fig. \ref{unitcellp}(b)) where an enormous confined magnetic field appears in the interior 
of the loop at resonance, when driven by an external source.

The size of the loop does not strongly influence the frequency of the individual magnetic resonance 
occurs. The plasmonic resonance
frequency of each nanoparticle primarily determines the resonant behaviour of the loop. 
However, the number of 
plasmonic particles on the loop determines the strength of the resonance~\cite{engheta_oe}. 
At resonance, the size of each unit cell ($a \sim \lambda/5$) and each loop inclusion is 
much smaller compared to the wavelength of the incident light ($R \sim \lambda/25$). Thus, 
Bragg scattering does 
not play a dominant role in the observed phenomenon and the system is easily homogenized.

\begin{figure}[tbp]
\includegraphics[angle= -0, width=1.0\columnwidth]{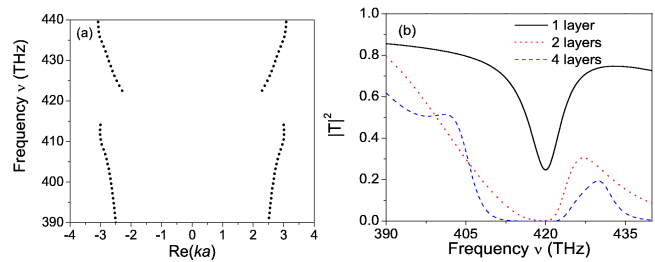}
\caption{(a) The band structure and (b), the transmittivity for four layers
of unit cells of the plasmonic metamaterial. The additional peaks 
are the Fabry-P\'{e}rot resonances occuring due to multiple
scattering between the interface of the medium and vacuum.
\label{bareplots}}
\end{figure}

The response of an array of such plasmonic loops was calculated using the PHOTON codes.
The calculations are essentially two dimensional, the cylindrical axis being the axis of 
invariance. To avoid staircasing effects with a square grid, we considered rods with square 
cross-sections rather than circular ones. The unit cell is a square with side $a = 150$nm 
(See Fig. \ref{unitcellp}(a)). Within each unit cell of the 
metamaterial, four silver nanorods (of square cross-section and side $b = 25$nm ),  
embedded in a background dielectric medium with $\epsilon_\mathrm{b} = 2.65$, are 
symmetrically arranged on the circumference of a circle of radius $\mathrm{R} = 26$ nm.
The metamaterial exhibits a distinctly identifiable magnetic resonance at 
$\sim$ 415 THz, with a negative magnetic 
permeability band gap up to $\sim$ 421 THz. This behaviour is also confirmed by the transmittivity
 for a slab  composed of four layers of unit cells (loops) 
(Fig. \ref{bareplots}(b)). We note that the resonance frequency of the 
metamaterial differs from that of the individual loop inclusions due to coupling 
between the nanoparticles. For smaller unit cell sizes and larger loop 
radii, the coupling between neighbouring plasmonic loop resonances leads to the 
hybridization of bands. Many weak resonances are found to occur close together and identification of the magnetic 
resonance leading to the negative permeability band gap, is complex for larger loop radii.    

\begin{figure}[tbp]
\includegraphics[angle= -0, width=1.0\columnwidth]{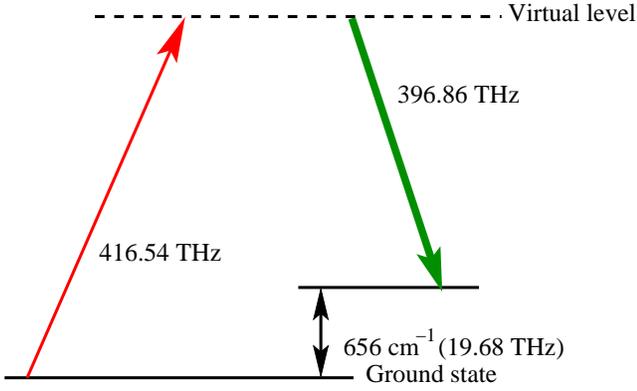}
\caption{The proposed level scheme for Inverse Raman Effect in $\mathrm{CS_2}$.\label{rlevels}}
\end{figure}

It is well-known that the plasmonic resonance of a metallic nanoparticle is highly
sensitive to the dielectric permittivity of the surrounding media ($\epsilon_b$). 
The plasmonic resonance of a cylindrical nanoparticle occurs when the condition
\begin{eqnarray}
\mathrm{Re}[ \epsilon_b + \epsilon_m] = 0,
\end{eqnarray}
is satisfied~\cite{novotny_hecht}. Using a Drude form for
 the permittivity of the  metal ($\epsilon_m$), the plasmon resonance frequency
 of a metallic cylinder may be written as:
 \begin{eqnarray}
 \omega_r = \omega_p \left[ \frac{1}{\epsilon_b + \epsilon_{\infty}} - \frac{\gamma^2}{\omega_p^2}\right]^{1/2}
 \end{eqnarray} 
 where $\omega_r$ is the plasmon resonance frequency for the nanoparticle, 
 $\omega_p$ and $\gamma$ defines the bulk plasma frequency 
 and the bulk dissipation of the metal, while $\epsilon_{\infty}$ is 
 its `static' permittivity. Hence, even a small change in
 $\epsilon_b$  can shift the plasmon resonance frequency of the nanorod and by extension, 
the magnetic loop resonance frequency arising out of the circulating displacement current 
around the loop.We note a subtle difference between the SRR and the plasmonic metamaterial
which may affect the efficiency with which their magnetic response is manipulated. In the case of 
the SRR, the displacement current is confined to the region within the capacitive gaps of the SRR. 
In this case, however, the displacement current is between adjacent nanoparticles, and hence, 
spatially delocalized. A more closely packed loop would, in principle, enable a higher degree of 
localization of currents and a stronger resonant response.

 Control of  $\epsilon_b$ by an applied electromagnetic
 radiation is the key to the control of the composite metamaterial itself. 
In order to demonstrate parametric control over the magnetic resonance, an atomic/molecular 
system exhibiting a resonance in the neighbourhood of the band gap frequency is essential.
Raman transitions that provide such resonances can be effectively used in a very flexible 
manner because the Raman resonance frequency can be tuned by the choice of the
pump laser frequency. A strong pump field with frequency $\nu_0$ will create conditions
 for the resonant absorption of a probe field at a frequency $(\nu_0 + \nu_R)$ (Inverse Raman Effect \cite{inv_raman1, inv_raman2}), where $\nu_R$ is the molecular level spacing of the medium.
We present  the calculated results for such a stimulated Raman absorption process 
(Fig.~\ref{rlevels}) at 416.54 THz 
in $\mathrm{CS_2}$ with ground state vibrational level spacing of $656 ~\mathrm{cm^{-1}}$ 
(19.68 THz). Experimentally obtained values for the parameters of $\mathrm{CS_2}$ were used 
\cite{yariv}, corresponding to the non-resonant $\chi^{(3)}$ response. We use peak pump powers of the order of 
100 $\mathrm{GW/cm^{2}}$ and a typical linewidth of $0.5 \mathrm{cm^{-1}}$(15 GHz). Our estimates for the nonlinearity 
are overtly conservative as the resonant Raman process would be expected to enhance the $\chi^{(3)}$ response. Such inverse 
Raman absorption of about -15dB has been reported in amorphous silicon by with much lower pump powers of about 
1 $\mathrm{GW/cm^{2}}$~\cite{solli_pra}, assuming a pump laser field at 396.86 THz which lies within the propagating band of the metamaterial (Fig.~\ref{bareplots}(a)).

\begin{figure}[tbp]
\includegraphics[angle= -0, width= 1.0\columnwidth]{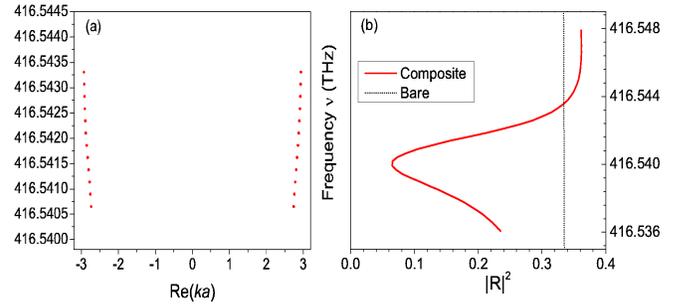}
\caption{(a) \& (b) The new band formed and the modified reflectivity of the 
plasmonic metamaterial in the neighbourhood
of a Raman transition in $\mathrm{CS_2}$ at 416.54 THz. The vertical line in (b) represents the reflectivity 
of the bare `metamaterial'.\label{pplots}}
\end{figure}

An additional propagating band develops within the band gap region seen in Fig. \ref{bareplots}, in the presence 
of the pump field 
and this propagating band region has been expanded in Fig.~\ref{pplots} 
This is manifested by a sharp drop of ($\sim$ 20\%) in the reflectivity of the medium. 
 The new band formed is quite dissipative and this prevents a large change in the 
transmittivity in comparison with the change in reflectivity. The transmittivity of the composite metamaterial is 
reasonable when compared to 
that for the `bare' metamaterial, but is completely overshadowed by the large change
in the reflectivity.
It is possible to numerically demonstrate dramatic enhancement of 
the transmittivity by several orders of magnitude
by using higher pump intensities. 
Thus, the effect of the introduction of the resonant Raman medium 
results in a sharp reduction in the reflectivity of the composite 
metamaterial over a narrow frequency band ($\sim 5$ GHz). 
It is the linewidth of the Raman pump
laser that dictates the linewidth of the induced absorption, which in turn determines 
the change in the reflectivity. The number of propagating bands that are formed are strongly dependent on the 
resonant nature of the background medium. In the case of EIT-based
control, additional bands which switch rapidly between states of low and high 
transmittivity, are formed in the presence of an control field $\Omega_c$~\cite{oe_iitk}.

The behaviour of the metamaterial for radiation of the other state of polarization 
(with the electric field of the incident radiation directed along the axis of the metallic rods) 
may also be controlled in a similar manner~\cite{OL_paper}. Fig.~\ref{unitcellp} (a) continues to represent 
a typical section of the metamaterial, 
considered here, however, with the $\mathrm{\vec{E}}$ field along the nanorod axes. 
As shown in Ref.~\cite{OL_paper}, the transmittivity of the effective plasma is zero while 
the reflectivity is high below the cut-off frequency. At higher frequencies, the 
transmittivity of the medium increases considerably. These correspond to the regions where 
$\epsilon_\mathrm{eff} < 0$ and $\epsilon_\mathrm{eff} > 0$, respectively. 
The response of this metamaterial can be actively manipulated 
 using embedded atomic/molecular media. The introduction of an  
embedding medium with a resonant response below the effective plasma frequency, into the 
metamaterial, will drive the effective permittivity to positive values over narrow frequency bands. 
This results in the formation of transmittive bands below $\omega_p$. 
Resonant enhancement of the background permittivity can also be achieved by 
stimulated Raman absorption due to an applied pump radiation in a similar manner. A new transmission band is found to develop at this frequency, which lies below the 
effective plasma frequency for the medium~\cite{OL_paper}. We further note that a lower linewidth of the pump laser results in a significant increase in transmittivity.
The plasma-like behaviour of
 the nanorod array depends on a bulk averaging process to determine its overall response and 
 does not necessitate a `loop'-like arrangement.
Thus, the response of the nanorod array to light of the two orthogonal states of linear polarization and 
their subsequent modified behaviour in the presence of a dispersive permittivity has
 a different origin.

\section{CONCLUSION}

We have thus demonstrated a new scheme for the control 
of metamaterials where the resonant metamaterial response is itself completely transformed 
by the introduction of controllable resonant media into the design. 
Coupled with the fact that metamaterials can be designed to show a resonant 
response at a predetermined frequency, this technique provides an entirely 
new way of reducing the dissipation in metamaterials, as our 
results show. We have discussed the control of SRR-based metamaterials by the parametric control of the 
capacitance, the control of nanorod plasmonic loop metamaterials by the control of the surface plasmon resonance via 
the background dielectric environment and the control of the plasma-like behaviour of the nanorod metamaterial by changing the bulk average dielectric permittivity.  The variety of such possibilities of control demonstrate the generality of this approach. The 
control scheme presented here enhances the tunability of metamaterials and can be effectively 
used for a variety of narrow-band applications, thus moving towards dynamically controllable
metamaterials.

\end{document}